\documentclass[12pt]{article}
\usepackage{amsmath,amsfonts}
\makeatletter \@addtoreset{equation}{section}

\usepackage{cite}
\usepackage{bm}
\usepackage{dcolumn}
\makeatother
\def\one{{\hbox{ 1\kern-.8mm l}}}
\newcommand{\Dslash}{\not{\hbox{\kern-4pt $D$}}}
\newcommand{\pdslash}{\not{\hbox{\kern-2pt $\partial$}}}
\newcommand{\be}{\begin{equation}}
\newcommand{\bea}{\begin{eqnarray}}
\newcommand{\eea}{\end{eqnarray}}
\newcommand{\ba}{\begin{array}}
\newcommand{\ea}{\end{array}}
\newcommand{\ee}{\end{equation}}
\newcommand{\nn}{\nonumber}

\begin{document}

\begin{titlepage}
  \thispagestyle{empty}

\vspace*{1mm}%
\hfill%
\vbox{
    \halign{#\hfil        \cr
           IPM/P-2010/024 \cr
                     } 
      }  
\vspace*{15mm}%


  \begin{center}
    \font\titlerm=cmr10 scaled\magstep4
    \font\titlei=cmmi10 scaled\magstep4
    \font\titleis=cmmi7 scaled\magstep4

     \centerline{\titlerm   Quasi-Normal Modes of Extremal BTZ Black Holes  in TMG}

\vspace*{15mm} \vspace*{1mm} {Hamid R. Afshar$^{a,b}$, Mohsen Alishahiha$^{a}$ and  Amir E. Mosaffa$^b$}

 \vspace*{1cm}

{\it ${}^a$ School of physics, Institute for Research in Fundamental Sciences (IPM)\\
P.O. Box 19395-5531, Tehran, Iran \\ }

\vspace*{.4cm}

{\it ${}^b$ Department of Physics, Sharif University of Technology \\
P.O. Box 11365-9161, Tehran, Iran}

\vspace*{2cm}

e-mails:\, afshar@ipm.ir,\, alishah@ipm.ir and mosaffa@physics.sharif.edu

\end{center}

\vspace*{2cm}

\begin{abstract}
We study the spectrum of tensor perturbations on extremal BTZ black holes in topologically massive gravity
for arbitrary values of the coefficient of the Chern-Simons term, $\mu$. Imposing proper boundary conditions
at the boundary of the space and at the horizon, we find that the spectrum contains quasi-normal modes.

\end{abstract}

\end{titlepage}

 \section{Introduction}
Topologically Massive Gravity (TMG) \cite{Deser:1982vy,Deser:1981wh} in three dimensions has
attracted a lot of attention recently owing to its rich structure.
Its wide range of solutions and also its interesting properties at
special values of its coupling constant has opened several
directions and venues for studying quantum gravity in the framework
of a toy model.

Recall that Einstein gravity in three dimensions, even with a
negative cosmological constant, is too trivial a theory at a first
glance. This is because the constraint equations are so strict that
do not allow any local propagating degrees of freedom. Yet, the
theory contains the famous BTZ black holes \cite{Banados:1992wn} as well as the
Brown-Henneaux boundary gravitons \cite{Brown:1986nw}. This theory is conjectured to be
dual to a, yet to be found, conformal field theory that lives on the
boundary of the space. The most trivial solution of the theory,
$AdS_3$ vacuum, is conjectured to be dual to the classical limit of
a highest weight state in the CFT whereas the boundary gravitons are
taken in correspondence with the descendant states that constitute a
Verma module on the highest weight.

Probably the most challenging question in this context is to give a
clear account for the microscopic origin of the BTZ black holes'
entropy. Perhaps the closest answer to this question has been given
in \cite{Witten:2007kt} where the BTZ black holes are interpreted as the
classical limit of highest weight states in the CFT. This CFT is
conjectured to have a global symmetry group that acts on the highest
weight states and the dimension of the representation determines the
microscopic degeneracy of the black hole states\footnote{For
some issues related to the quantum behaviour of there dimensional gravity,
see {\it e.g.} \cite{{Orlando:2010ay},{Israel:2004vv},{Detournay:2005fz}}.}.

Propagating degrees of freedom can appear if higher derivative
corrections are added to the above system. In TMG these corrections
are provided by the gravitational Chern-Simons action
\bea\label{MG}
I_{TMG}&=&\frac{1}{16\pi G}\left(I_{EH}+\frac{1}{\mu}I_{CS}\right),
\eea
with
\bea
I_{EH}=\int d^3 x \sqrt{-g} \left(R+\frac{2}{\ell^2}\right),\;\;\;\;
I_{CS}=\frac{1}{2}\int d^3
x\sqrt{-g}\epsilon^{\lambda\mu\nu}\Gamma^{\rho}_{\lambda\sigma}\left(\partial_{\mu}
\Gamma^{\rho}_{\lambda\sigma}+\frac{2}{3}\Gamma^{\sigma}_{\mu\tau}\Gamma^{\tau}_{\nu\rho}\right),
\eea
where $\ell$ determines the cosmological constant through
$\Lambda=-1/\ell^2$ and $\mu$ is a coupling constant.

TMG contains all the solutions of Einstein gravity including the
$AdS_3$ vacuum. Linearized equations of motion around this vacuum
first showed the appearance of a local propagating massive normal
mode $\psi^M$ as well as the usual left and right moving boundary
gravitons $\psi^L$ and $\psi^R$ \cite{Li:2008dq}. It was then shown that either the massive
mode or the BTZ black holes have a negative mass unless the coupling
constant takes the value $\mu=1/\ell$ \cite{Li:2008dq}. At this point $\psi^M$ and
$\psi^L$ become degenerate with zero energy and can be removed. This is
because exactly at this point gauge symmetry is enhanced to include
all the left moving Virasoro generators. What remains is the tower of right
moving boundary gravitons which are massless and of course the whole
spectrum of BTZ black holes that have non negative masses. The point
$\mu=1/\ell$ was thus called the chiral point and it was conjectured that
at this point we are left with a chiral unitary theory.

Soon after, it was shown in \cite{Grumiller:2008qz} that the linearized equations of motion
at the critical point have a solution which may be interpreted as a left moving
excitation\footnote{Whether at the critical point the model is really chiral or not has been further investigated in several
papers including \cite{{Carlip:2008jk},{Park:2008yy},{Grumiller:2008pr},{Carlip:2008eq},{Carlip:2008qh},
{Giribet:2008bw},{Blagojevic:2008bn},{Li:2008yz},{Maloney:2009ck}}. See also \cite{Balasubramanian:2009bg}
for a rigorous definition of chiral CFT. }.
However it is worth mentioning that this new mode which has the same asymptotic behavior as
AdS wave solutions \cite{{AyonBeato:2004fq},{AyonBeato:2005qq}} does not obey Brown-Henneaux
boundary conditions \cite{Brown:1986nw}. Therefore if we restrict ourselves to solutions
which satisfy the Brown-Henneaux conditions one may still have a chiral theory at least classically. On the other hand
if we relax the boundary conditions the theory will not be chiral and indeed it was
conjectured in \cite{Grumiller:2008qz} and proved in \cite{{Skenderis:2009nt},{Grumiller:2009mw}} that the dual theory (in the sense of AdS/CFT correspondence \cite{Maldacena:1997re})
could be a logarithmic CFT (LCFT) \cite{Gurarie:1993}\footnote{Such a behavior has also been
appeared in NMG \cite{{Bergshoeff:2009hq},{AyonBeato:2009yq},{Grumiller:2009sn},{Alishahiha:2010bw}}, Born-Infeld
gravity \cite{{Gullu:2010pc},{Nam:2010dd},{Alishahiha:2010iq}} as well as bigravity \cite{Afshar:2009rg}.}.

To explore some features of TMG we study quasi-normal modes (QNM's) of the tensor perturbations in the
model. Having put  the subject in the context of AdS/CFT correspondence \cite{Maldacena:1997re}, the study of QNM's
of asymptotically AdS black holes becomes more interesting as they
are giving us information about the behavior of the dual thermal CFT
that is living on the boundary of the space time \cite{{Kovtun:2004de},{Kovtun:2005ev}}. For usual BTZ
black holes in a theory of pure gravity, gravitational tensor QNM's cannot
exist as there are no local propagating degrees of freedom in the
bulk. Therefore in such theories, the focus has only been on scalar,
fermion or vector perturbations on BTZ black holes
\cite{Birmingham:2001pj}\footnote{Quasi-normal modes of quantum corrected BTZ black hole have been
studied in \cite{Konoplya:2004ik}. See also \cite{Cardoso:2001hn}
for early  discussions on BTZ perturbations and quasi-normal modes.}.

However, in TMG, as well as other higher derivative three
dimensional gravities, gravitons can propagate and hence
gravitational QNM's become relevant. Such an analysis has recently
been done for non-extremal BTZ black holes in TMG
\cite{{Sachs:2008gt},{Sachs:2008yi}}
(for scalar perturbation in TMG
see \cite{Lee:2008gta}). In the following we extend these results to
the case of extremal BTZ black holes in TMG which requires an
independent analysis and which demonstrate some exclusive behaviors (for scalar and fermion perturbation on extremal BTZ see
\cite{Crisostomo:2004hj}).
For a recent review on QNM's see\cite{Berti:2009kk}.

The paper is organized as follows. In section two we will analyze  gravitons
on extremal BTZ black holes in TMG by making use of the linearized equations of motion.
Using the result of section two we shall study the QNM's of the tensor perturbations in section three.
the last section is devoted to discussions.

\section{Gravitons on extremal BTZ black hole}

In this section we make a detailed analysis of gravitons on extremal
BTZ black holes in TMG for arbitrary values of $\mu \ell$.
We choose the Gaussian Normal coordinate \cite{Carlip:2005zn,Banados:1998gg} in which
the BTZ black hole is given by (for more information see appendix A)
\bea\label{normalcoord}
ds^2=\ell^2\bigg[L^+du^2+L^-dv^2+d\rho^2
-\left(e^{2\rho}+L^+L^-e^{-2\rho}\right)dudv\bigg],
\eea
where $u=t/\ell-\phi$, $v=t/\ell+\phi$ and
\bea
L^\pm=\frac{(r_+\pm r_-)^2}{4\ell^2}
\eea
To get the $AdS_3$ solution one needs to set $r_+^2=-1$ and $r_-=0$, while the extremal BTZ black
hole is given in the limit of $r_+=r_-$ where the corresponding metric in the Gaussian Normal coordinate becomes \cite{{Crisostomo:2004hj},{Balasubramanian:2009bg}}
\bea\label{extbtz}
ds^2=\bar{g}_{\mu\nu}dx^{\mu}dx^{\nu}=r_{ex}^2du^2-\ell^2e^{2{\rho}}du
dv+\ell^2d{\rho}^2
\eea
Our aim is to study the spectrum of linear perturbations around this background.
To proceed, consider the following perturbations
\bea
g_{\mu\nu}=\bar{g}_{\mu\nu}+h_{\mu\nu}.
\eea
Since the background is stationary and static one may start with a general ansatz for
the excitations, $\psi$, as follows (see however \cite{Compere:2010xu})
\bea\label{wave}
\psi_{\mu\nu}(u,v,\rho)=e^{-ihu-i\bar{h}v}F_{\mu\nu}(\rho),
\eea
whose real part represents the excitations of the metric; ${\rm Re} (\psi_{\mu\nu})=h_{\mu\nu}$.
It should also be noted that since the direction $\phi$ is periodic, the corresponding momentum along
this direction must be quantized. In other words one has
\bea
h-\bar{h}=k\in\mathbb{Z}.
\eea
By making use of this ansatz the linearized equations of motion reduce to two first order
coupled differential equations\footnote{From now on we set $\ell=1$.}
\bea\label{difLeom}
F'_{vv}=-(\mu-1)F_{vv}-i\bar{h}F_{\rho v},\;\;\;\;\;\;\;\;
F'_{\rho v}=-(\mu+1)F_{\rho v}-i\bar{h}F_{\rho \rho},
\eea
together with the following four algebraic equations\footnote{This is consistent with the fact that the massive graviton in
$D$-dimensional spacetime has $ \frac{(D+1)(D-2)}{2} $ degrees of freedom.}
\bea\label{algLeom}
(\mu+1)F_{\rho u}e^{2\rho}&=&2i(hF_{uv}-\bar{h}F_{uu}),\;\;\;\;\;\;\;
\mu F_{\rho\rho}e^{4\rho}=2i(hF_{\rho v}-\bar{h}F_{\rho u})e^{2\rho}+4F_{vv}r_{ex}^2,\cr &&\\
(\mu-1)F_{\rho v}e^{2\rho}&=&2i(hF_{vv}-\bar{h}F_{uv}),\;\;\;\;\;\;\;\;
F_{\rho\rho}e^{4\rho}=4(r_{ex}^2F_{vv}+e^{2\rho}F_{uv}),\nonumber
\eea
From these equations it is straightforward to find that $F_{vv}$ satisfies the following
second order differential equation
\bea\label{diffeq}
F_{vv}''+2F_{vv}'+\left[4(\bar{h}h+\bar{h}^2r_{ex}^2e^{-2\rho})e^{-2\rho}-(\mu-1)(\mu-3)\right]F_{vv}=0.
\eea
It is evident from the above differential equation that for the case of $\bar{h}=0$ the equation has
no propagating solutions and therefore we will assume that $\bar{h}\neq0$. In this case it is useful to define
a new variable\footnote{The asymmetry of equations
under $h\leftrightarrow\bar{h}$ is a consequence of the absence of
$vv$ component in the background metric.}
\bea\label{zdef}
 z=2i\bar{h}r_{ex}e^{-2\rho}
\eea
by which the equation (\ref{diffeq}) can be recast to a familiar equation,  Whittaker equation, as follows
\bea
\frac{d^2F_{vv}}{dz^2}+\left[-\frac{1}{4}+\frac{\lambda}{z}+\frac{\frac{1}{4}-m^2}{z^2}\right]F_{vv}=0,
\eea
where
$\lambda=\frac{h}{2ir_{ex}}$ and $m=\pm(\frac{\mu}{2}-1)$. Since the equation is symmetric under the
sign of $m$, in the rest of the paper we only consider the case of $m=\frac{\mu}{2}-1$.

The most general solution of the above differential equation which is suitable for all ranges of
$m$\cite{book} is
\bea\label{persol}
F_{vv}=C_1W_{\lambda,m}(z)+C_2W_{-\lambda,m}(-z),
\eea
where
\bea
W_{\lambda,m}(z)=\frac{\Gamma(-2m)}{\Gamma(\frac{1}{2}-m-\lambda)}M_{\lambda,m}(z)
+\frac{\Gamma(2m)}{\Gamma(\frac{1}{2}+m-\lambda)}M_{\lambda,-m}(z),
\eea
and
\bea
M_{\lambda,m}(z)=z^{m+1/2}e^{-z/2}{}_1F_1(m-\lambda+\frac{1}{2},1+2m;z),
\eea
with ${}_1F_1$ being the Kummer confluent hypergeometric function.

Having found $F_{vv}$, the other components can be obtained from equations
\eqref{algLeom} and \eqref{difLeom} as follows
\bea\label{othercmpts}
F_{v\rho}&=&\frac{i}{\bar{h}}\left[(\mu-1)F_{vv}+F'_{vv}\right]\cr &&\cr
F_{uv}&=&-\frac{1}{2\bar{h}^2}\left[((\mu-1)^2e^{2\rho}-2h\bar{h})F_{vv}+(\mu-1)e^{2\rho}F'_{vv}\right]\cr &&\cr
F_{\rho\rho}&=&-\frac{2}{\bar{h}^2}\left[((\mu-1)^2-2\bar{h}^2r_{ex}^2e^{-4\rho}-2h\bar{h}e^{-2\rho})F_{vv}+(\mu-1)F'_{vv}\right]\cr &&\cr
F_{u\rho}&=&-\frac{i}{\bar{h}^3}\left[(\mu(\mu-1)^2e^{2\rho}-2\bar{h}^2r_{ex}^2(\mu-1)e^{-2\rho}+h\bar{h}(1-3\mu))F_{vv}+(\mu(\mu-1)e^{2\rho}-h\bar{h})F'_{vv}\right]\cr &&\cr
F_{uu}&=&\frac{1}{2\bar{h}^4}\big[\left(\mu(\mu+1)(\mu-1)^2e^{4\rho}-4\mu^2
h\bar{h}e^{2\rho}+2(h^2-r_{ex}^2(\mu^2-1))\bar{h}^2\right)F_{vv}\cr &&\cr &&\;\;\;\;\;\;\;+\mu\left((\mu^2-1)e^{2\rho}-2h\bar{h}\right)e^{2\rho}F_{vv}'\big]
\eea

\section{Quasi-Normal Modes }

Black holes as thermodynamic systems, can be studied under small
perturbations. The decay of these perturbations are described by
quasi-normal modes. In fact the relaxation time for the decay of the
black hole perturbation is determined by the imaginary part of the
lowest quasi-normal mode. In this section we would like to study the QNM in
the extremal black hole.

\subsection{Direct derivation}

In order to determine QNM's, one needs to solve the wave equation in
the black hole background with the specific boundary conditions
at horizon and  the conformal boundary.
Usually one assumes that the wave function vanishes at the boundary while it should be
an ingoing wave at the horizon. The appearance of QNM's is the reflection of the fact that these
boundary conditions lead to frequencies with a non-zero imaginary part.
To see whether we have QNM's for the extremal back hole in TMG one needs to impose the
above boundary conditions on the solution that we have found in the previous section.

By looking at the behavior of the Whittaker's function at $z\rightarrow \infty$ (at horizon)
\bea
W_{\lambda,m}(z)\sim e^{-z/2}z^{\lambda},\;\;\;\;\;\;\;\;\;W_{-\lambda,m}(-z)\sim e^{z/2}z^{-\lambda},
\eea
and from the equation (\ref{persol}) one can see that at the horizon one may have two
different modes as follows
\bea
\psi_{vv}(t,\rho)\sim e^{-i(h+\bar{h})t}W_{\pm\lambda,m}(\pm z)\sim
e^{-i[(h+\bar{h})t\pm \bar{h}r_{ex}e^{-2\rho}]\mp2\lambda \rho}.
\eea
It is then evident that for ${\rm Re}(1+h/\bar{h})>0$, the solution (\ref{persol}) gives an ingoing wave at the horizon provided $C_1=0$. As we will see, since $h$ is purely imaginary, the condition ${\rm Re}(1+h/\bar{h})>0$
is always satisfied. On the other hand taking into account that
\bea
F'_{vv}\sim e^{-2\rho}F_{vv}\;\;\;\;\;\;\text{and}\;\;\;\;\;\;F_{vv}\sim e^{i\bar{h}r_{ex}e^{-2\rho}}\;\;\;\;\;\text{as}\;\;\;\;\; \rho\rightarrow-\infty,
\eea
and from the equation \eqref{othercmpts}, we find that if $F_{vv}$ is ingoing at the
horizon, then all other components will be ingoing as well.

Having imposed the condition at the horizon one needs to be sure that the solution vanishes as we
approach the boundary. Therefore we should impose the condition that the dominant component
of the solution \eqref{othercmpts} is zero at the boundary, $z=0$.

To proceed it is worth noting that the near boundary behavior of the Whittaker's function, where $z\rightarrow 0$,
is given by \cite{book}
\bea \label{near}
W_{-\lambda,m}(z)\sim
\frac{\Gamma(-2m)}{\Gamma(\frac{1}{2}-m+\lambda)}\;z^{\frac{1}{2}+m}
+\frac{\Gamma(2m)}{\Gamma(\frac{1}{2}+m+\lambda)}\;z^{\frac{1}{2}-m}.
\eea
Therefore, whatever the dominant component would be, the possible poles one might have could come
from either the condition   $\frac{1}{2}+m+\lambda=0,-1,\cdots$ or $\frac{1}{2}-m+\lambda=0,-1,\cdots$.

\subsubsection*{Case 1}
Consider the case where the poles are are given by $\frac{1}{2}+m+\lambda=-n,$ in which
\be
h=-ir_{ex}(2n+\mu-1),\;\;\;\;\;\;\;{\rm for}\;\;n=0,1,2,\cdots.
\ee
In this case the dominant component at the boundary will be $F_{vv}$ (see appendix B) which has the following near boundary
behavior
\bea
F_{vv}\sim
\frac{\Gamma(-2m)}{\Gamma(\frac{1}{2}-m+\lambda)}\;z^{\frac{1}{2}+m}.
\eea
We thus find that this component vanishes as $z\rightarrow 0$, if $m+\frac{1}{2}>0$ or $\mu>1$. Note also that the lowest mode ($n=0$) also
decays in time if $\mu>1$. Therefore in this case we get non-trivial QNM's for $\mu>1$ whose frequencies are
given by
\be\label{QNM}
\omega_n=k-2ir_{ex}(2n+\mu-1),\;\;\;\;\;\;\;\;{\rm for}\;\;n=0,1,2,\cdots.
\ee

\subsubsection*{Case 2}
If $\frac{1}{2}-m+\lambda=-n$, for which
\be
h=-ir_{ex}(2n+3-\mu),\;\;\;\;\;\;\;{\rm for}\;\;n=0,1,2,\cdots,
\ee
it is evident form \eqref{othercmpts}, and also from our discussions in the appendix B, that the
dominant component is $F_{uu}$. The behavior of this component at
the boundary is now given by
\bea
F_{uu}\sim e^{4\rho}F_{vv}\sim
\frac{\Gamma(2m)}{\Gamma(\frac{1}{2}+m+\lambda)}\;z^{-\frac{3}{2}-m},
\eea
which vanishes for $m+\frac{3}{2}<0$ or $\mu<-1$. On the other hand for the lowest mode to be a decaying mode in
time one needs to have $\mu<3$ which is automatically satisfied. So for $\mu<-1$ we get QNM's whose frequencies are given by
\be\label{QNM1}
\omega_n=k-2ir_{ex}(2n+3-\mu),\;\;\;\;\;\;\;\;{\rm for}\;\;n=0,1,2,\cdots.
\ee

To summarize,  we find QNM's for  tensor perturbations
on  extremal BTZ black holes in TMG when $\mu l < -1$ or $\mu l > 1$. The cases of $\mu l =\pm 1$ will be
discussed later. This result is particularly interesting because QNM's are usually associated with thermal properties of the horizon. For an extremal
horizon, a priori, it is not
obvious why we should still have QNM's. Before making speculations on this point, we would like to compare
our results with those in the literature by rederiving QNM's using a certain procedure that has recently been used in \cite{Sachs:2008gt},
namely, by imposing the so called \textit{chiral highest weight condition}.
This is the subject of next subsection.

\subsection{Chiral highest weight condition}

In this subsection we would like to compare our procedure to those in the
literature (see \cite{Sachs:2008gt}). The main observation is that
the QNM's we have found in the previous subsection are given in terms of the
Whittaker's functions which satisfy a certain
recursive relation\cite{book}
\bea
z\frac{d}{dz}W_{\lambda,m}(z)=\left(\lambda-\frac{z}{2}\right)W_{\lambda,m}(z)-\left[m^2-\big(\lambda-\frac{1}{2}\big)^2\right]W_{\lambda-1,m}(z).
\eea
In terms of the $\rho$ coordinate, this relation may be recast to the following form
\be
\frac{1}{2r_{ex}}\bigg(-r_{ex}\partial_\rho-ih-i\bar{h}r_{ex}^2e^{-2\rho}\bigg)W_{-\lambda,m}(-z)
=\left(\frac{1}{2}+m+\lambda\right)\left(\frac{1}{2}-m+\lambda\right)\;W_{-\lambda-1,m}(-z).
\ee
Restricting our attention to the lowest mode, in which either  $\frac{1}{2}+m+\lambda$ or
$\frac{1}{2}-m+\lambda$ is zero, and taking into account that the $u$ and $v$ dependence of the
wave function is given by \eqref{wave}, the above equation reduces to
\be\label{chrlhgst}
-\frac{1}{2r_{ex}} \bigg(\partial_u+2r_{ex}^2e^{-2\rho}-r_{ex}\partial_\rho\bigg)e^{-ihu-i\bar{h}v}F_{vv}=0.
\ee
Now consider the following operators
\bea\label{gener}
  L_0&=&-\frac{1}{2r_{ex}}\partial_{u},\cr &&\cr
  L_1&=&-\frac{e^{2r_{ex}u}}{2r_{ex}}[\partial_{u}+2r_{ex}^2e^{-2\rho}\partial_{v}-r_{ex}\partial_{\rho}], \cr &&\cr
  L_{-1}&=&-\frac{e^{-2r_{ex}u}}{2r_{ex}}[\partial_{u}+2r_{ex}^2e^{-2\rho}\partial_{v}+r_{ex}\partial_{\rho}].
\eea
These operators satisfy the  $SL(2,R)$ algebra and are identified with the Killing vectors of the
background \eqref{extbtz}. We therefore find out that the recursive relations for Whittaker functions require our lowest QNM to be  annihilated by $L_1$. That is,
the lowest mode satisfies a chiral highest weight condition.

Motivated by the above observation, one may turn around the argument as follows;  start by imposing the chiral highest weight
condition on the lowest mode, solve the equations of motion and then impose the proper
boundary conditions. That is, we start with

\be
L_1{\psi^{(0)}}_{\mu\nu}=0.
\ee
Using the notation of the previous subsection the above condition can be solved easily,
\bea
F_{vv}&=&E_1z^{-\lambda}e^{z/2},\cr &&\cr
F_{v\rho}&=&(E_2-\frac{iE_1}{\bar{h}}z)z^{-\lambda}e^{z/2},\cr &&\cr
F_{\rho\rho}&=&(E_3-2\frac{iE_2}{\bar{h}}z-\frac{E_1}{\bar{h}^2}z^2)z^{-\lambda}e^{z/2},\cr &&\cr
F_{uv}&=&r_{ex}(2i\bar{h}E_4z^{-1}+E_2)z^{-\lambda}e^{z/2},\cr &&\cr
F_{u\rho}&=&r_{ex}(\frac{E_2}{i\bar{h}}z+(E_3+2E_4)+2i\bar{h}E_5z^{-1})z^{-\lambda}e^{z/2},\cr &&\cr
F_{uu}&=&r_{ex}^2(E_3+4i\bar{h}E_5z^{-1}-4\bar{h}^2E_6z^{-2})z^{-\lambda}e^{z/2},
\eea
where $E_i$'s are constants to be determined by the equations of motion. In fact from the linearized equations of
motion, $D^M\psi_{\mu\nu}=0$, one finds
\bea
E_2&=&\frac{h+ir_{ex}(\mu-1)}{\bar{h}r_{ex}}E_1,\cr &&\cr
E_3&=&2\frac{(h+ir_{ex}(\mu-1))(h+2ir_{ex})}{(\bar{h}r_{ex})^2}E_1,\cr &&\cr
E_4&=&\frac{(h+ir_{ex}(\mu-1))(h+2ir_{ex})}{2(\bar{h}r_{ex})^2}E_1,\cr &&\cr
E_5&=&\frac{(h+ir_{ex}(\mu-1))(h+2ir_{ex})(h+3ir_{ex})}{(\bar{h}r_{ex})^3}E_1,\cr &&\cr
E_6&=&\frac{(h+ir_{ex}(\mu-1))(h+2ir_{ex})(h+3ir_{ex})(h+4ir_{ex})}{(\bar{h}r_{ex})^4}E_1.
\eea
Moreover the highest weight, $h$, is found to be either $h=-ir_{ex}(\mu-1)$ or $h=ir_{ex}(\mu-3)$, which are indeed
the poles of the lowest mode obtained in the previous subsection.
In particular for the case of $h=-ir_{ex}(\mu-1)$ the solution reads
\bea
F_{\mu\nu}=e^{-i(\frac{h}{r_{ex}}\rho-\bar{h}r_{ex}e^{-2\rho})}\left(
  \begin{array}{ccc}
    0 & 0 & 0 \\
    0 & 1 & 2r_{ex}e^{-2\rho} \\
    0 & 2r_{ex}e^{-2\rho} & 4r_{ex}^2e^{-4\rho} \\
  \end{array}
\right),
\eea
which has to be compared with \eqref{S1} obtained from the direct computations. This represents a QNM if
$\mu>1$ with the frequency $\omega=k-2ir_{ex}(\mu-1)$, in agreement with \eqref{QNM}.

On the other hand for the other case, $h=ir_{ex}(\mu-3)$, the solution is given by
\bea
E_2&=&\frac{2i}{\bar{h}}(\mu-2)E_1,\cr &&\cr
E_3&=&\frac{-4}{\bar{h}^2}(\mu-2)(\mu-1)E_1,\cr &&\cr
E_4&=&\frac{-1}{\bar{h}^2}(\mu-2)(\mu-1)E_1,\cr &&\cr
E_5&=&\frac{-2i}{\bar{h}^3}\mu(\mu-2)(\mu-1)E_1,\cr &&\cr
E_6&=&\frac{1}{\bar{h}^4}\mu(\mu-2)(\mu^2-1)E_1,
\eea
which represents a QNM if $\mu<-1$ with the frequency $\omega=k-2ir(3-\mu)$ in agreement with \eqref{QNM1}.

Note that what we have found in this subsection is the lowest QNM in two different cases given by
the highest weights $h=-ir_{ex}(\mu-1)$ or $h=ir_{ex}(\mu-3)$. This confirms our calculations in the previous section.

It is important to also note that imposing the chiral highest weight condition guarantees that the the solutions we find are ingoing at the horizon.
Moreover from the above considerations one would expect
that the higher QNM's can be obtained by apply $L_{-1}$ on the lowest mode. This is indeed the case if we use  another
recursive relation,
\bea\label{recus}
W_{-\lambda+1,m}(-z)=(-1/2z+\lambda)W_{\lambda,m}(-z)-z\frac{d}{dz}W_{-\lambda,m}(-z).
\eea
Writing the $n$th mode as
\bea\
{\psi^{(n)}}_{\mu\nu}(u,v,\rho)=e^{-ih_nu-i\bar{h}_nv}W_{-\lambda_n,m}(-z),
\eea
where $\bar{h}_n=h_n-k$, $h_n=2ir_{ex}\lambda_n$ and $\lambda_n=-\left(\frac{1}{2}\pm m+n\right)$, the $(n+1)$th mode is
found by using \eqref{recus} and the fact that two consecutive $\lambda$'s are related by $\lambda_{n+1}=\lambda_n-1$,
\bea
{\psi^{(n+1)}}_{\mu\nu}(u,v,\rho)=e^{-2r_{ex}u-2r_{ex}v}[-z/2+\lambda_n-z\partial_z]{\psi^{(n)}}_{\mu\nu}(u,v,\rho).
\eea
This can be written in a more suggestive form as
\bea
{\psi^{(n+1)}}_{\mu\nu}(u,v,\rho)=-e^{-2r_{ex}v}L_{-1}{\psi^{(n)}}_{\mu\nu}(u,v,\rho),
\eea
where $L_{-1}$ is given in (\ref{gener}). We therefore find that
\bea
{\psi^{(n)}}_{\mu\nu}(u,v,\rho)=\left(-e^{-2r_{ex}v}L_{-1}\right)^n{\psi^{(0)}}_{\mu\nu}(u,v,\rho).
\eea
Comparing this with the non-extremal case, the factor of  $-e^{-2r_{ex}v}$ is all that remains
from $\bar{L}_{-1}$.

\subsection{QNM's for $\mu=\pm 1$}

We now turn to the question of whether QNM's persist to exist at the special values of the coupling constant,  $\mu=\pm1$.

Looking at the solutions we found in the previous sections at the special value of , say, $\mu=1$, we find that they do not
fall off at the boundary, and besides, the lowest frequency has no imaginary part. This might lead to the conclusion that we no longer have QNM's at this point.
But we should note that exactly at this point,  the solutions we have
found degenerate with the left moving boundary gravitons and are no longer propagating. However,
a new propagating mode appears at this point which is logarithmic and which is given by
\bea
\psi_{\mu\nu}^{new}&=&\frac{d\psi_{\mu\nu}}{d\mu}\big{|}_{\mu=1}
=-i\left(u\frac{dh}{d\mu}+v\frac{d\bar{h}}{d\mu}\right)_{\mu=1}\psi_{\mu\nu}+e^{-ihu-i\bar{h}v}
F_{\mu\nu}^{new},
\eea
where $F^{new}_{\mu\nu}=\frac{dF_{\mu\nu}}{d\mu}\big{|}_{\mu=1}$. At $\mu=1$ with $h=-2ir_{ex}n$, where
the dominant component is $F_{vv}$, one finds
\bea
\psi_{vv}^{new}=\left(-2r_{ex}tF_{vv}+F_{vv}^{new}\right)e^{-ik(t-\phi)-4nr_{ex}t}.
\eea
Using the explicit expressions for $F_{vv}$ one observes that for $n=0$ as we approach the boundary
one gets
\be
\psi^{new}_{\mu\nu}\rightarrow \infty,
\ee
whereas for $n\geq 1$ we have
\be
\psi^{new}_{\mu\nu}\rightarrow {\rm finite}.
\ee
Therefore we get no QNM's at $\mu=1$.

At $\mu=-1$ with $h=-2ir_{ex}(2+n)$, where the dominant component is $F_{uu}$, we have
\bea
\psi_{uu}^{new}=\left(2r_{ex}tF_{uu}+F_{uu}^{new}\right)e^{-ik(t-\phi)-4(n+2)r_{ex}t}.
\eea
Plugging the solutions in this expression we find that  we have QNM's in this case even for
$n=0$ with frequencies given by \eqref{QNM1} with $\mu=-1$.

\subsection{Standing waves}
To complete our analysis of the tensor perturbations, we will consider the case of
$\bar{h}= 0$. Note that the results we have found so far were based on the assumption
that $\bar{h}\neq 0$ (see equation \eqref{zdef}). Setting $\bar{h}= 0$ one encounters two
possibilities depending on whether $h$ is zero or not. Of course in both cases, the perturbations result in
standing waves.

In the first case where $h\neq 0$ and  $\mu\neq-1,-2$, the equations \eqref{difLeom} and \eqref{algLeom} can be solved
to give
\bea\label{rr}
F_{vv}(\rho)&=&C_1(\mu-1)e^{-(\mu-1)\rho}\cr &&\cr
F_{\rho v}(\rho)&=&2ihC_1e^{-(\mu+1)\rho}\cr &&\cr
F_{\rho\rho}(\rho)&=&\frac{4C_1}{\mu}[(\mu-1)r_{ex}^2-h^2]e^{-(\mu+3)\rho}\cr &&\cr
F_{uv}(\rho)&=&-\frac{C_1}{\mu}[(\mu-1)^2r_{ex}^2+h^2]e^{-(\mu+1)\rho}\\
F_{\rho u}(\rho)&=&-\frac{2ihC_1}{\mu(\mu+1)}[(\mu-1)^2r_{ex}^2+h^2]e^{-(\mu+3)\rho}\cr &&\cr
F_{uu}(\rho)&=&\frac{C_1}{\mu(\mu+1)(\mu+2)}[(\mu+1)^2r_{ex}^2+h^2][(\mu-1)^2r_{ex}^2+h^2]e^{-(\mu+3)\rho}+C_2e^{(\mu+1)\rho}.\nonumber
\eea
As we anticipated, the above solution is not well defined for $\mu=-1$ and $\mu=-2$. Indeed for the
latter case where  $F_{uu}$ is singular, one finds that the solution is modified to one
with a logarithmic behavior. More precisely in this case  $F_{uu}$ is given by
\bea
F_{uu}(\rho)=[-C_1(h^2+9r_{ex}^2)(h^2+r_{ex}^2)\rho+C_2]e^{-\rho}.
\eea
The other components can be read off from \eqref{rr} with $\mu=-2$. For the case of  $\mu=-1$, the equations
\eqref{difLeom} and \eqref{algLeom} put a constrain on the parameters
$h^2+4r_{ex}^2=0$. The corresponding solution is
\bea
F_{uv}=0,\;\;F_{vv}=C_1e^{2\rho},\;\;F_{\rho\rho}=4C_1r_{ex}^2e^{-2\rho},
\;\;F_{\rho v}=-iC_1h,\;\;F_{\rho u}=C_2e^{-2\rho},\;\;F_{uu}=C_3+\frac{ihC_2}{2}e^{-2\rho}.\nn\\
\eea

On the other hand for $\bar{h}=h=0$, from \eqref{difLeom} and \eqref{algLeom} we observe that
$\mu$ must be either $\mu=1$ or $\mu=-1$. For the former case one finds
\be
F_{\rho u}=F_{uv}=0,\;\;\;F_{vv}=C_1,\;\;\;F_{\rho\rho}=4C_1r_{ex}^2e^{-4\rho},
\;\;\;F_{\rho v}=C_2e^{-2\rho},\;\;\;F_{uu}=C_3e^{2\rho}.
\ee
while for $\mu=-1$ we get
\bea
F_{\rho v}=0,\;\;F_{\rho u}=C_1e^{-2\rho},\;\;F_{vv}=C_2e^{2\rho},\;\;
F_{uv}=-2r_{ex}^2C_2,\;\;F_{\rho\rho}=-4C_2r_{ex}^2e^{-2\rho},\;\;F_{uu}=C_3.\nn\\
\eea
We note, however, that at $\mu=\pm 1$, the model develops logarithmic solutions. In our notation we get
\bea
&&{\rm for}\;\;\mu=1,\;\;\;\;\;F^{new}_{uu}=\frac{d F_{uu}}{d\mu}|_{\mu=1}\sim \rho e^{2\rho},\cr &&\cr
&&{\rm for}\;\;\mu=-1,\;\;\;\;\;F^{new}_{uu}=\frac{d F_{uu}}{d\mu}|_{\mu=-1}\sim \rho ,
\eea
which are the logarithmic solutions found in \cite{Grumiller:2008qz}.

\section{Discussions}

In this paper we have considered tensor perturbations on the extremal BTZ black hole in TMG. We have found that,
imposing proper boundary conditions at the boundary of the space and at the horizon, one finds QNM's in the spectrum.
These modes vanish for $r_{ex}=0$, where
we have $M=J=0$, or $T_L=T_R=0$ from the boundary CFT point of view.

We note that our considerations do not have a counterpart in pure Einstein gravity in three dimensions.
This is due to the fact the gravity in three dimensions even with a negative cosmological constant
does not have propagating degrees of freedom. Higher derivative terms allow for such propagating modes.

Since QNM's are usually associated with thermal properties of the black holes, it is
quite interesting that the model we are considering supports QNM's even for extremal
black holes. This is indeed one of the special features of TMG.

Recall that for generic values of $\mu$, TMG contains negative masses either in
perturbative states or in the spectrum of BTZ black holes. Even
at the chiral points $\mu=\pm1$, the appearance of logarithmic modes render the system non-unitary
unless a strict Brown-Hennaux boundary condition is imposed. Whether or not such an assumption remains valid
at a quantum level, allowing for a chiral unitary sector to decouple at the chiral points, remains to be verified. A natural venue
for such an analysis is the AdS/CFT correspondence where one can study the $n$-point functions of the dual field theory
and see whether the above decoupling is possible. Another direction is to calculate the gravitational
partition function of TMG around any of its vacua.

The non-unitarity of TMG might explain the appearance of QNM's on extremal black holes. Since we have allowed
logarithmic boundary conditions for perturbations, these modes persist to exist at the chiral point.
It would be interesting to
understand the consequences of the result from the dual CFT.

As a final remark we recall that the boundary condition we have imposed was to have an
ingoing wave as we approach the horizon and all of our results rely on this assumption.
Another possibility is to impose a vanishing flux at the horizon and at the boundary. Using the properties of the Whittaker's function,
one can see that this assumption requires $\bar{h}=0$ which, as we have
seen,  results in standing modes.

\section*{Acknowledgments}

We would like to thank Ali Naseh for collaboration in the early stages of this project. We would also like to thank O. Saremi and M. M. Sheikh-Jabbari for useful discussions
and also A. Akhvan, A. Davody, R. Fareghbal and A. Vahedi for discussions on different aspects of NMG and TMG.

\section*{Appendix}

\appendix

\section{ Gaussian Normal Coordinates}
The metric of BTZ black holes is usually written as
\bea\label{BTZBH}
ds^2=-N^2dt^2+\frac{dr^2}{N^2}+r^2(d\phi+N^{\phi}dt)^2,
 \eea
  with
\bea
N^2=-8MG+\frac{r^2}{\ell^2}+\frac{16G^2J^2}{r^2}\;\;\;\;\;\;\;N^{\phi}=-\frac{4GJ}{r^2}.
\eea
 If we define
 \bea
M=\frac{r_+^2+r_-^2}{8G\ell^2}\,,\;\;\;\;\;\;J=\frac{r_+r_-}{4G\ell},
\eea
 then the metric takes the form
 \bea
ds^2=-\frac{(r^2-r_+^2)(r^2-r_-^2)}{r^2\ell^2}dt^2+\frac{r^2\ell^2}{(r^2-r_+^2)(r^2-r_-^2)}dr^2+r^2(d\phi-\frac{r_+r_-}{r^2\ell}dt)^2.
\eea
By the following change of coordinates
\bea
r^2=r_+^2\cosh^2(\rho-\rho_0)-r_-^2\sinh^2(\rho-\rho_0),\;\;\;\;\;\;\;\;\;e^{2\rho_0}=\frac{r_+^2-r_-^2}{4\ell^2},
\eea
we go to the Gaussian Normal coordinates
\cite{Banados:1998gg,Carlip:2005zn}
\bea
ds^2=\ell^2[L^+du^2+L^-dv^2+d\rho^2
-\left(e^{2\rho}+L^+L^-e^{-2\rho}\right)dudv],
\eea
where
$u=t/\ell-\phi$, $v=t/\ell+\phi$ and
\bea
L^\pm=\frac{(r_+\pm
r_-)^2}{4\ell^2}.
\eea
Changing the variables as
\bea\label{coshads}
\frac{e^{2\rho}}{\sqrt{L^+L^-}}=e^{2\hat{\rho}},
\eea
we obtain
\bea
ds^2=\ell^2[L^+du^2+L^-dv^2+d\hat{\rho}^2
-2\sqrt{L^+L^-}\cosh(2\hat{\rho})dudv].
\eea
Pure $\text{AdS}_3$ is obtained by setting
\bea r_+^2=-1,\;\;\;\;\;\;\;r_-=0.
\eea
If we make a second coordinate transformation as
\bea
\hat{u}=2u\sqrt{L^+},\;\;\;\;\;\;\;\;\hat{v}=2v\sqrt{L^-},
\eea
we have
\bea
ds^2=\ell^2\left[\frac{1}{4}d\hat{u}^2+\frac{1}{4}d\hat{v}^2+d\hat{\rho}^2-\frac{1}{2}\cosh(2\hat{\rho})d\hat{u}d\hat{v}\right],
\eea
which is the BTZ black hole with unit mass and zero angular
momentum.

In the extremal limit, $r_+=r_-=r_{ex}$, the transformation \eqref{coshads} is not well defined. Starting from
\bea
ds^2=-\frac{(r^2-r_{ex}^2)^2}{r^2\ell^2}dt^2+\frac{r^2\ell^2}{(r^2-r_{ex}^2)^2}dr^2+r^2(d\phi-\frac{r_{ex}^2}{r^2\ell}dt)^2,
\eea
the transformation
\bea
e^{2\rho}=\frac{r^2-r_{ex}^2}{\ell^2},
\eea
takes us to
\bea\label{ExtBTZ}
ds^2=\bar{g}_{\mu\nu}dx^{\mu}dx^{\nu}=r_{ex}^2du^2-\ell^2e^{2{\rho}}du
dv+\ell^2d{\rho}^2,
\eea
where $u$ and $v$ are defined as before. In this coordinate
$\rho=-\infty$ corresponds to the location of horizon,
$r=r_{ex}$, and $\rho=\infty$ corresponds to $r=\infty$.

\section{Dominant component}

In this appendix we show that in the first case where the poles are given by $\frac{1}{2}+m+\lambda=0,-1,-2,\cdots,$
the dominant component is $F_{vv}$, as we approach the boundary. To proceed we note that the
Whittaker's function obeys the following recursive relation \cite{book}
\bea
z\frac{d}{dz}W_{\lambda,m}(z)=\left(\lambda-\frac{z}{2}\right)W_{\lambda,m}(z)-\left[m^2-\big(\lambda-\frac{1}{2}\big)^2\right]W_{\lambda-1,m}(z).
\eea
Therefore setting
\bea\label{pole3}
\frac{1}{2}+m+\lambda=-n\;\;\;\; \text{with}\;\;\;\;n=0,1,2,\cdots,
\eea
one gets
\bea\label{dFvv1}
F'_{vv}=(2\lambda-z)C_2W_{-\lambda,m}(-z)+2n\left(\lambda-m+\frac{1}{2}\right)C_2W_{-\lambda-1,m}(-z).
\eea
On the other hand, at the poles \eqref{pole3}, by making use of the asymptotic behavior of the
Whittaker's function one finds
\bea\label{asFvv}
F_{vv}&=&C_2W_{-\lambda,m}(-z)=C_2\frac{\Gamma(-2m)}{\Gamma(\frac{1}{2}-m+\lambda)}(-z)^{\frac{1}{2}+m}
\bigg(1-\frac{\lambda}{2m+1}z+\mathcal{O}(z^2)\bigg),\\ &&\cr
F'_{vv}&=&C_2\frac{\Gamma(-2m)}{\Gamma(\frac{1}{2}-m+\lambda)}(-z)^{\frac{1}{2}+m}\left(2(\lambda+n)-\frac{2\lambda^2+2n(\lambda+1)+2m+1}{2m+1}z+\mathcal{O}(z^2)\right),\nonumber
\eea
as $z\rightarrow 0$. Plugging these expressions in \eqref{othercmpts}
we arrive at
\bea
F_{uv}&=&-\frac{1}{\bar{h}}\bigg[ir_{ex}(2m+1)(2\lambda+2m+1+2n)z^{-1}+\mathcal{O}(1)\bigg]F_{vv},\nn\\
F_{u\rho}&=&-\frac{2i}{\bar{h}^2}\bigg[ir_{ex}(2m+2)(2m+1)(2\lambda+2m+1+2n)z^{-1}+\mathcal{O}(1)\bigg]F_{vv},\nn\\
F_{uu}&=&-\frac{2}{\bar{h}^2}\bigg[r_{ex}^2(2m+3)(2m+2)(2m+1)(2\lambda+2m+1+2n)z^{-2}\nn\\
&&\;\;\;\;\;\;\;\;-2r_{ex}^2(2m+2)(2m\lambda+2m+5\lambda+3)(2\lambda+2m+1+2n)z^{-1}+\mathcal{O}(1)\bigg]F_{vv},\nn
\eea
which shows that the leading terms evaluated at the poles \eqref{pole3} are zero. Therefore all components, at most,
could be in the same order as $F_{vv}$. As a result to find the QNM's it is enough to impose the boundary condition
on $F_{vv}$.

It is worth mentioning that since we are evaluating the expressions at the poles \eqref{pole3}, one has
\bea
\frac{\Gamma(-2m)}{\Gamma(\frac{1}{2}-m+\lambda)}=(-1)^n(2m+1)_n\;\;\;\; \text{where}\;\;\;\;(a)_n=a(a+1)(a+2)\cdots(a+n),
\eea
which shows that at the points $\mu=2-k$, $k=1,2,\cdots,n$, the dominant term in the above expansion starts from $\mathcal{O}(z^{m+\frac{1}{2}+k})$.
Nevertheless since we are interested in the case where $\mu>1$, these points
do not affect the validity of our conclusion.

Note that for the lowest mode, where $h=-ir_{ex}(\mu-1)$, we get
\bea\label{othercmpts1}
F_{uv}&=&-\frac{1}{2ir_{ex}\bar{h}^2}\left(h+ir_{ex}(\mu-1)\right)\left[(\mu-1)e^{2\rho}-2i\bar{h}r_{ex}\right]
F_{vv},\cr &&\cr
F_{u\rho}&=&-\frac{1}{r_{ex}\bar{h}^3}\left(h+ir_{ex}(\mu-1)\right)\left[\mu(\mu-1)e^{2\rho}-\bar{h}(h-2\mu ir_{ex})-2\bar{h}^2r_{ex}^2e^{-2\rho}\right]F_{vv},\cr &&\cr
F_{uu}&=&\frac{1}{2ir_{ex}\bar{h}^4}\left(h+ir_{ex}(\mu-1)\right)\big[\mu(\mu^2-1)e^{4\rho}-2\bar{h}(h+ir_{ex}(\mu+1))(\mu e^{2\rho}+\bar{h}r_{ex})\big]F_{vv},\nn
\eea
which are all zero. This leads to the following expression for the perturbations
\bea \label{S1}
F_{\mu\nu}=F_{vv}\left(
  \begin{array}{ccc}
    0 & 0 & 0 \\
    0 & 1 & 2r_{ex}e^{-2\rho} \\
    0 & 2r_{ex}e^{-2\rho} & 4r_{ex}^2e^{-4\rho} \\
  \end{array}
\right).
\eea

\end{document}